\documentclass[pre,aps,twocolumn,showpacs]{revtex4}
\usepackage{epsfig}
\usepackage{graphicx}

\begin{document}
\title{Spectral formulation and WKB approximation for rare-event statistics in reaction
systems}
\author{Michael Assaf and Baruch Meerson}
\affiliation{Racah Institute of Physics, Hebrew University of Jerusalem,
Jerusalem 91904, Israel} \pacs{05.40.-a, 02.50.Ey, 82.20.-w, 87.23.Cc, 03.65.Sq}

\begin{abstract}
We develop a spectral formulation and a stationary WKB approximation for
calculating the probabilities of rare events (large deviations from the mean) in
systems of reacting particles with infinite-range interaction, describable by a
master equation. We compare the stationary WKB approximation with a recent
\textit{time-dependent} semiclassical approximation developed, for the same
class of problems, by Elgart and Kamenev. As a benchmark we use an exactly
solvable problem of the binary annihilation reaction $2A \to \emptyset$.
\end{abstract}
\maketitle
\section{Introduction}
Since the pioneering works of Delbr\"{u}ck \cite{delb}, Bartholomay \cite{barth}
and McQuarrie \textit{et al.} \cite{McQuarrie1,McQuarrie}, kinetics of reacting
systems with infinite-range interaction, containing a large but finite number of
molecules or agents (such as bacteria, cells, animals or even humans) have
attracted much attention \cite{gardiner,vankampen}. While the change in time of
the average number of particles in such systems  may be describable by
(continuum) rate equations, one often needs to know the probability of a
non-typical behavior. This necessitates going beyond the rate equations, and a
standard way of achieving this goal is provided by the \textit{master equation}
of a gain-loss type which directly deals with $P_{n_1,n_2,...,n_N}(t)$: the
probability of simultaneously having $n_{1}$ particles of the first type,
$n_{2}$ particles of the second type, $\dots$, and $n_{N}$ particles of the
$N$th type at time $t$ \cite{gardiner,vankampen}. Though providing a complete
description, the master equation is rarely solvable analytically, so various
approximations are in use \cite{gardiner,vankampen}. Probably the most widely
used approximation is the Fokker-Planck equation which usually suffices when one
is \textit{not} interested in extreme statistics, such as extinction time
\cite{kamenev,Sander}. Being interested in extreme statistics, we will exploit
here a well-known mathematical formulation (see, \textit{e.g.} \cite{gardiner})
which, though exactly equivalent to the master equation, deals instead with a
generating function $G(\mathbf{x},t)$ which encodes all the probabilities
$P_{n_1,n_2,...,n_N}(t)$. The generating function is defined in the following
way:
\begin{equation}\label{genfunmany}
G(\mathbf{x},t) = \sum_{n_{1}=0,\,\dots,\,
n_{N}=0}^{\infty}\left(\prod_{i=1}^{N}x_{i}^{n_{i}}\right)P_{n_1,n_2,...,n_N
}(t)\,,
\end{equation}
whereas the probabilities $P_{n_1,n_2,...,n_N}(t)$ are recovered by
differentiation:
\begin{equation}\label{probmany} \left.P_{n_1,\dots,n_N}(t) =
\frac{1}{\prod_{i=1}^{N}n_{i}!}\frac{\partial^{n_1+\dots+n_N}G(\mathbf{x},t)}{\partial
x_{1}^{n_1}\cdots\partial x_{N}^{n_N}} \right|_{\mathbf{x}=\mathbf{0}}\,.
\end{equation}
This formulation transforms the master equation (an infinite set of differential
difference equations) into a single linear partial differential equation for
$G(\mathbf{x},t)$:
\begin{equation}\label{geneqmany}
\frac{\partial G(\mathbf{x},t)}{\partial t}= \hat{L} G(\mathbf{x},t)\,,
\end{equation}
where $\hat{L}$ is a real linear differential operator which involves
derivatives with respect to the auxiliary variables $\mathbf{x}=(x_{1}, x_{2},
\dots , x_{N})$. The initial condition for this equation is supplied by Eq.
(\ref{genfunmany}) with the time-dependent probabilities replaced by their
(prescribed) values at $t=0$. When the rate constants are time-independent, the
operator $\hat{L}$ is independent of time. There is one universal
\textit{boundary} condition in this formulation. Indeed, as
\begin{equation}\label{conservation}
    \sum_{n_{1}=0,\,\dots,\,
n_{N}=0}^{\infty} P_{n_1,n_2,...,n_N}(t) = 1\,,
\end{equation}
one immediately gets
\begin{equation}\label{universalmanyBC}
    G(x_1=1, \dots, x_N=1; t)=1\,.
\end{equation}
Correspondingly, $\hat{L} G(\mathbf{x},t)$ must vanish at $\mathbf{x}=(x_{1},
x_{2}, \dots , x_{N})=(1, 1, \dots, 1)$.

This paper will be limited to a single species, $N=1$. In this case the
generating function $G(x,t)$ is
\begin{equation}\label{generatingfun}
G(x,t)=\sum_{n=0}^{\infty}x^{n}P_{n}(t)\,,
\end{equation}
the probabilities $P_{n}(t)$ are
\begin{equation}\label{probformula}
\left.P_{n}(t)=\frac{1}{n!}\frac{\partial^{n}G(x,t)}{\partial
x^{n}}\right|_{x=0}\,,
\end{equation}
the evolution equation for $G$ is
\begin{equation}\label{generaleq}
\frac{\partial G(x,t)}{\partial t}= \hat{L} G(x,t)\,,
\end{equation}
and the universal boundary condition (\ref{universalmanyBC}) is
\begin{equation}\label{universalBC}
    G(x=1,t)=1\,.
\end{equation}
The initial condition is
\begin{equation}\label{incond}
    G(x,0) \equiv G_{0}(x)=\sum_{n=0}^{\infty}x^{n}P_{n}(0)\,.
\end{equation}

We will be interested in the important class of problems where the operator
$\hat{L}$ is of the \textit{second} order \cite{secondorder}. The crux of our
approach is that, for any time-independent second-order linear differential
operator $\hat{L}$, one can expand $G(x,t)$ in a complete set of properly
constructed orthogonal spatial eigenfunctions of the operator $\hat{L}$
\cite{hermit,Arfken}. The linear \textit{ordinary} differential equation for
these eigenfunctions can be obtained, for any specific problem, by separation of
variables. By a proper change of variables one can always eliminate the first
derivative from this equation, and arrive at a spectral problem for a
\textit{stationary} Schr\"{o}dinger equation for a zero-energy particle in a
potential $V(x,\mu)$ which depends on a parameter $\mu$ coming from the
separation of variables. This parameter, unknown \textit{a priori}, represents
the eigenvalue of this problem. This spectral formulation is exact, and it paves
the way to a systematic computation of the probabilities $P_n(t)$, where $n$ can
be significantly different from the ``typical", or average number of particles.
Using the probabilities, one can accurately estimate a host of quantities of
interest, for example, the average extinction time and the lifetime
distribution. The present work is mainly concerned with one useful technique
within the framework of the spectral formulation: the WKB approximation
\cite{Landau,orszag} which yields semiclassical eigenvalues and eigenfunctions.
Using these, one can construct an approximate solution of the initial value
problem for $G(x,t)$ and, by virtue of Eq. (\ref{probformula}), calculate
$P_n(t)$ for $n \gg 1$.

As Eq. (\ref{geneqmany}) is readily interpretable as a (non-Hermitian)
time-dependent Schr\"{o}dinger equation with imaginary time, there were earlier
quantum mechanical interpretations of rare event statistics in reacting systems,
such as the Doi-Peliti formalism of second quantization
\cite{Doi,Peliti,Mattis}. Still another approach to this class of problems has
been recently suggested by Elgart and Kamenev \cite{kamenev}. Instead of dealing
with the creation and annihilation operators, as is customary in the second
quantization approach, Elgart and Kamenev reformulated the time-dependent
problem in semiclassical terms, employing two strong inequalities: $n\gg 1$ and
$\bar{n}(t)\gg 1$, where $\bar{n}(t)$ is the average number of particles in the
system at time $t$. They showed that classical dynamics corresponding to the
Hamiltonian of the problem provide a valuable information about the rare-event
statistics, and that this approach is greatly superior to the more customary
Fokker-Planck description. Elgart and Kamenev start with the ansatz
$G(x,t)=\exp[-S(x,t)]$ in Eq. (\ref{generaleq}). Then, neglecting the
$\partial_{xx}S$ term, they arrive at a Hamilton-Jacobi equation for $S(x,t)$
which, for a time-independent $\hat{L}$, is solvable \cite{landaumech}. This
procedure yields $G(x,t)$ and, by virtue of Eq. (\ref{probformula}), the
probabilities $P_n(t)$. Elgart and Kamenev illustrated this approach on several
pedagogical examples which included various combinations of binary annihilation,
branching, decay and creation of particles.

As explained above, we suggest in this work a different type of semiclassical
approximation for this non-Hermitian quantum mechanics: a stationary WKB
approximation based on an exact spectral formulation. We will show that the two
semiclassical approximations complement each other, each of them being
advantageous in some region of the parameter space. We will demonstrate our
approach by a simple example of binary annihilation reaction $2A
\to\hspace{-4.3mm}^{\lambda}\hspace{2mm} \emptyset$, where $\lambda>0$ is the
rate constant. In this case the master equation is
\begin{equation}\label{masterann}
 \frac{d}{dt}{P}_{n}(t)=\frac{\lambda}{2}\left[(n+2)(n+1)
P_{n+2}(t)-n(n-1)P_{n}(t)\right]\,,
\end{equation}
while the evolution equation for $G(x,t)$ takes the form
\begin{equation}\label{anndiffeq}
\frac{\partial G}{\partial t} = \frac{\lambda}{2}(1-x^{2})\frac{\partial^{2}
G}{\partial x^{2}}\,.
\end{equation}
This example is instructive for two reasons. First, it is one of the examples
used by Elgart and Kamenev \cite{kamenev} to illustrate their time-dependent
semiclassical approach. Second, and no less important, it is exactly solvable.
McQuarrie \textit{et al.} \cite{McQuarrie} used the exact solution to find the
average number of particles and the variance versus time, when starting from a
fixed even number of particles. We significantly extend the analytical solution
in Appendix and find the probabilities $P_{n}(t)$ for all $n$ and $t$, and their
various asymptotics. Using these findings, we also calculate the probability
distribution of lifetimes of the particles in this system and the average
extinction time. These exact results provide a benchmark for our WKB theory. In
principle, the strong inequality $n \gg 1$ is the only criterion required in
this theory for \textit{all} times. We observed, however, that in practice one
also needs to require $n \gtrsim \bar{n}$, in order to avoid loss of accuracy
resulting from summation of many large terms of alternating sign. We will show
that, in the region of $n \gtrsim \bar{n}$, the stationary WKB formalism is much
more accurate than the time-dependent formalism due to Elgart and Kamenev, while
in the region of $n\lesssim \bar{n}$ the time-dependent formalism (which
circumvents the summation of large terms of alternating sign) is advantageous.

Here is a layout of the rest of the paper. In Section II we apply separation of
variables to Eq. (\ref{anndiffeq}), arrive at a Sturm-Liouville eigenvalue
problem and find approximate solutions to this problem by using the WKB
approximation. Then we solve, in Section III, an initial value problem, compute
the respective approximate probabilities $P_n(t)$ and compare them with the
exact probabilities and their asymptotics, derived in Appendix. In Section IV we
compare the predictions of the time-dependent semiclassical formulation
\cite{kamenev} with the exact results and establish the validity of the
time-dependent and stationary semiclassical approximations. Section V presents a
brief summary and discussion of our results.

\section{Spectral formulation and WKB approximation}
\subsection{Boundary conditions and steady state solution}
To complete the formulation of the problem for Eq. (\ref{anndiffeq}), we need
two boundary conditions. The first of them, Eq. (\ref{universalBC}), is
universal. The second one, at $x=-1$, readily follows from Eq. (\ref{anndiffeq})
itself: $G(x=-1,t)=C =const$, where $C=G_0(x=-1)$ is determined by the initial
data (\ref{incond}) \cite{second}.

The limit of $t \to \infty$ corresponds to a \textit{steady state} solution of
Eq. (\ref{anndiffeq}): $G_s(x)=A+Bx$. To obey the boundary conditions at $x=\pm
1$, we choose $A=(1+C)/2$ and $B=(1-C)/2$. When $C=1$ (an even number of
particles at $t=0$), the steady state solution $G_s(x)=1$ corresponds to an
empty system: $P_n=\delta_{n0}$, where $\delta_{ij}$ is the Kroenecker delta.
When $C=-1$ (an odd number of particles at $t=0$), one obtains $G_s(x)=x$. This
corresponds to a single particle, $P_n=\delta_{n 1}$, which lives forever as
there are no particles it can react with.

\subsection{Separation of variables and eigenvalue problem}
\begin{figure}
\includegraphics[width=8cm,clip=]{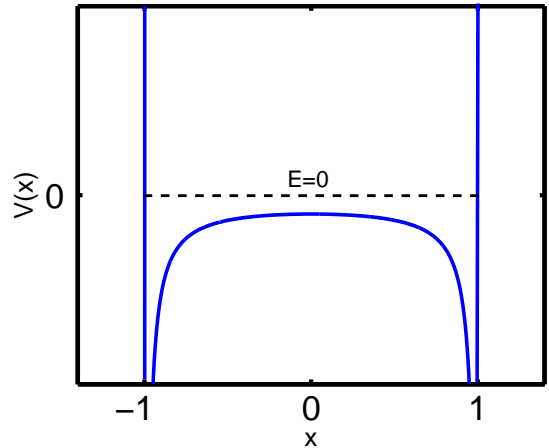}
\caption{(Color online) Shown by the blue solid lines is the singular potential
$V(x)$, given by Eq. (\ref{potential}). The vertical scale is arbitrary.}
\label{poten}
\end{figure}

Now let us consider the time-dependent part of the generating function:
$g(x,t)=G(x,t)-G_{s}(x)$. As $g(x,t)$ must vanish at $x=\pm 1$, we can look for
solutions  of the equation for $g(x,t)$,
\begin{equation}\label{homog}
   \frac{\partial g}{\partial t} =
\frac{\lambda}{2}(1-x^{2})\frac{\partial^{2} g}{\partial x^{2}}\,,
\end{equation}
in the separable form $g(x,t)=\exp(-\gamma t)\,\varphi(x)$. We obtain
\begin{equation}\label{oanndiffeq}
\varphi^{\prime\prime}(x)+\frac{\mu^2 \varphi(x)}{1-x^{2}}=0\,,
\end{equation}
where $\mu^{2}=2\gamma/\lambda$, and $\varphi(\pm 1)=0$. Equation
(\ref{oanndiffeq}) can be interpreted as a stationary Schr\"{o}dinger equation
for a zero-energy particle ($m=\hbar =1$) in the singular potential
\begin{equation}\label{potential}
    V(x)=\left\{\begin{array}{rcl}- \frac{\mu^2}{2(1-x^2)}\,,\;\;\;&&|x|<1\\
    +\infty\,,\;\;\;&&|x|\ge 1\,,
    \end{array} \right.
\end{equation}
see Fig. \ref{poten}, with (\textit{a priori} unknown) magnitude $\mu^2$ which
plays the role of eigenvalue. The problem is exactly solvable in terms of
Legendre polynomials, and the solution is presented in Appendix. In the next
subsection we will proceed as if we were unaware of the exact solution, and find
the spectrum of $\mu$ and the eigenfunctions in the WKB approximation.

\subsection{Stationary WKB approximation: wave functions and quantization}

\begin{figure}
\includegraphics[width=8cm,clip=]{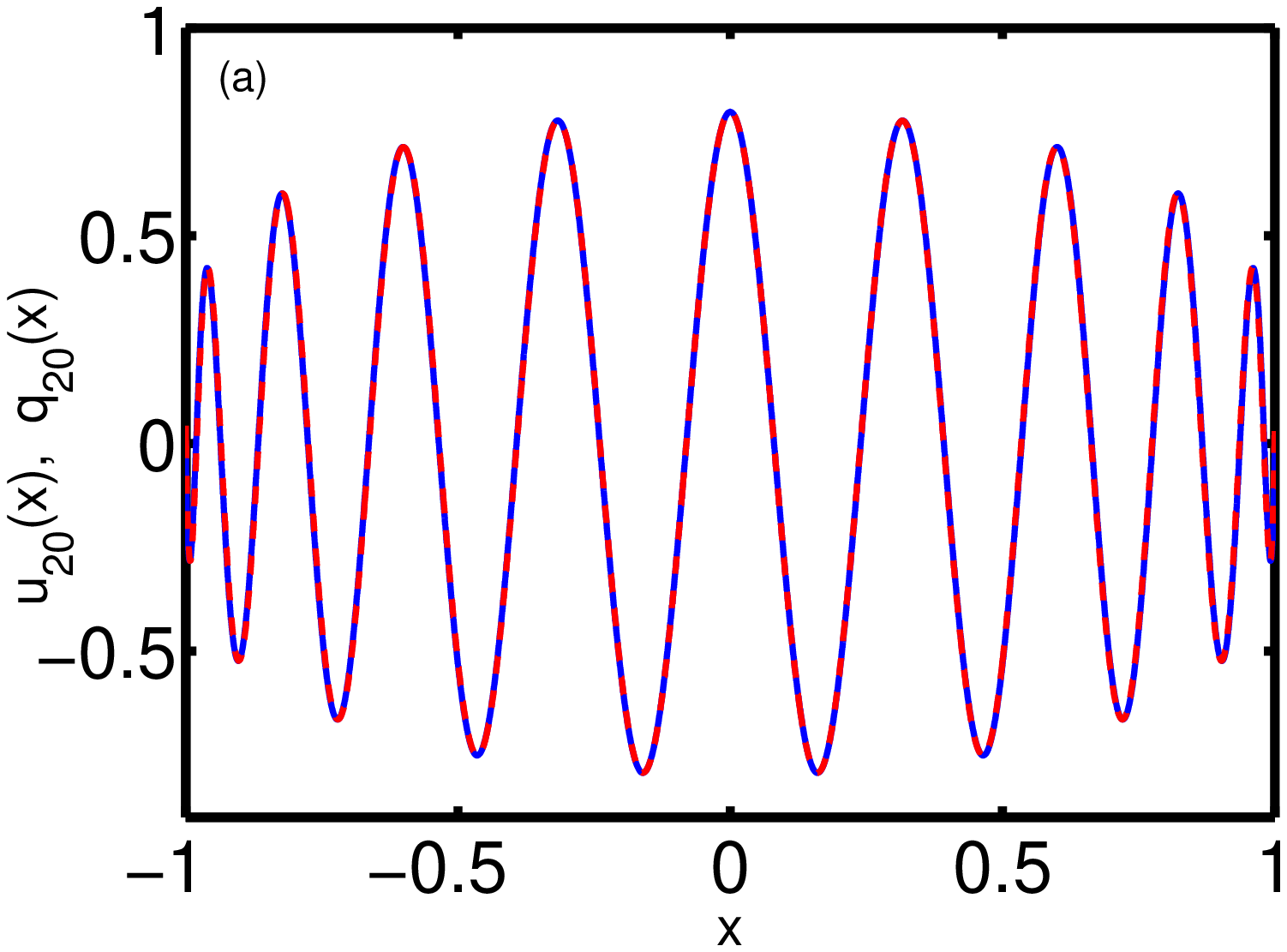}
\includegraphics[width=8cm,clip=]{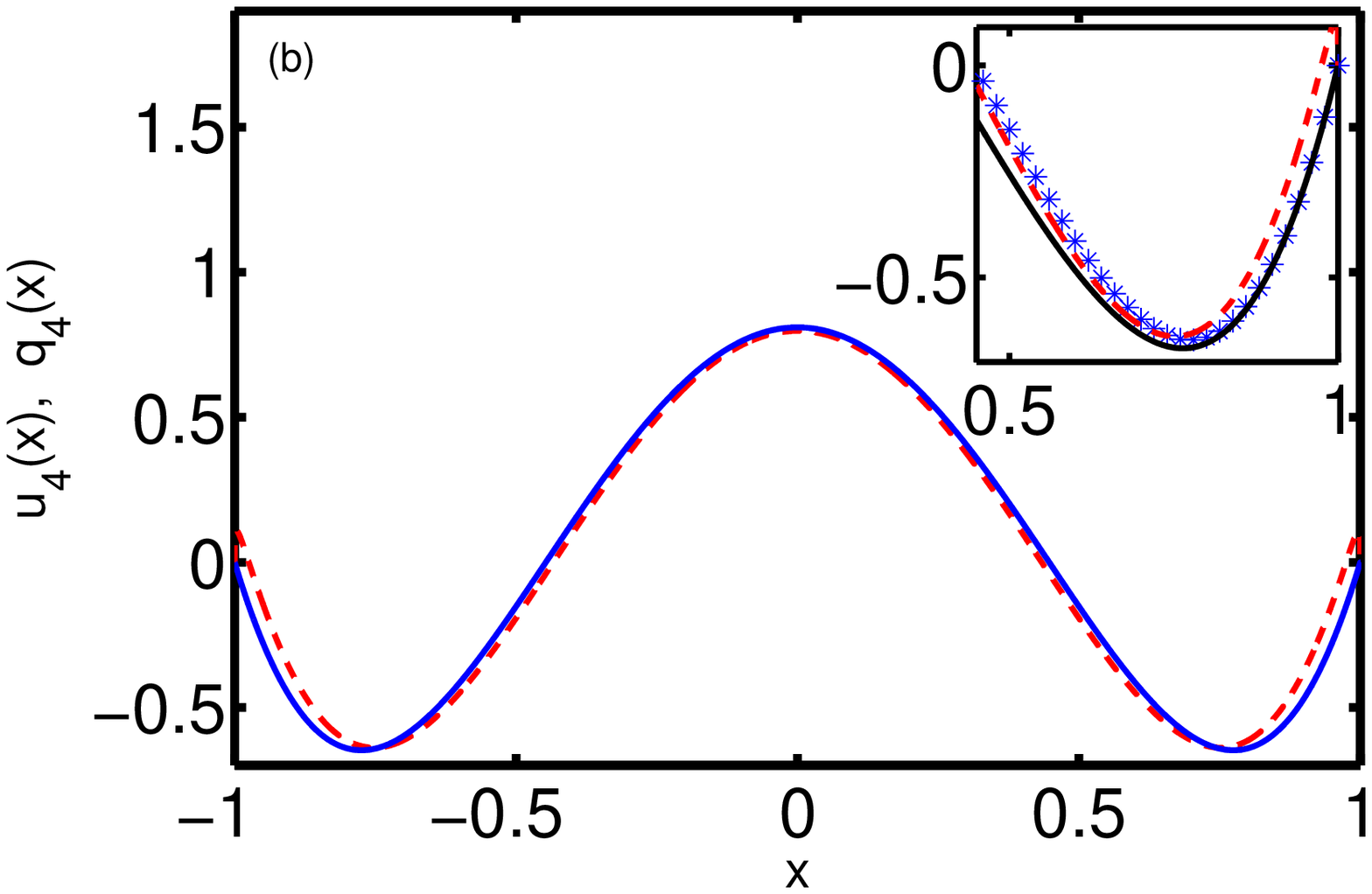}
\caption{(Color online) The normalized WKB eigenfunctions
(\ref{wkbeigenfunction}) (the red dashed lines) and the normalized exact
eigenfunctions (the blue solid lines) for $k=10$ (a) and $k=2$ (b). As one can
see, a good agreement is observed even for $k = 2$, while for $k=10$ the
agreement is excellent. The inset in (b) shows a close vicinity of $x=1$. One
can see that the function $u_{20}^{(a)}(x)$ (the red dashed line) deviates from
the exact solution (indicated by the crosses) in the vicinity of $x=1$, whereas
the function $u_{20}^{(b)}(x)$ (the solid line) is in excellent agreement with
the exact solution there.} \label{eigen}
\end{figure}

The crucial assumption of our semiclassical theory is that the main contribution
to the probabilities $P_n(t)$ with $n \gg 1$ comes from the
\textit{semiclassical} region of spectrum of $\mu$, where $\mu \gg 1$ and the
eigenfunctions have multiple zeros on the interval $|x|<1$. We will verify this
assumption \textit{a posteriori}. Employing the strong inequality $\mu \gg 1$,
we find by a standard calculation \cite{Landau,orszag} two independent (even and
odd) WKB solutions of Eq. (\ref{oanndiffeq}):
\begin{eqnarray}
\varphi_{even}(x)&\simeq&(1-x^{2})^{1/4}\cos(\mu\arcsin x)
\label{wkbsolann} \\
\varphi_{odd}(x) &\simeq&(1-x^{2})^{1/4}\sin(\mu\arcsin x)\,. \label{wkbodd}
\end{eqnarray}
To quantize the eigenvalue $\mu$, we must use the boundary conditions at $x=\pm
1$. The WKB solutions (\ref{wkbsolann}) and (\ref{wkbodd}) are invalid, however,
at and near the singular points $x=\pm 1$. To determine the solutions there we
first solve Eq. (\ref{oanndiffeq}) in a small vicinity of each of these points.
Here it suffices to consider the point $x=1$. Let us introduce a new coordinate
$\xi=1-x\ll{1}$ and neglect the subleading terms in the expansion of the
potential $V(x)$ near $x=1$. Equation (\ref{oanndiffeq}) becomes
\begin{equation}\label{end}
2\xi\varphi^{\prime\prime}(\xi)+\mu^2\varphi(\xi)=0\,.
\end{equation}
The two independent solutions are
\begin{equation}\label{first}
\varphi_{1}(\xi)=\xi^{1/2}J_{1}[\mu(2\xi)^{1/2}]
\end{equation}
and
\begin{equation}\label{second}
\varphi_{2}(\xi)=\xi^{1/2}Y_{1}[\mu(2\xi)^{1/2}]\,,
\end{equation}
where $J_{1}$ and $Y_{1}$ are the Bessel functions of the first and second kind,
respectively. Only $\varphi_{1}(\xi)$ obeys the required boundary condition
$\varphi(\xi=0)=0$, so $\varphi_{2}(\xi)$ must be discarded. Now we can match
$\varphi_1 (\xi)$ with each of the WKB solutions (\ref{wkbsolann}) and
(\ref{wkbodd}). Indeed, when $\mu \gg 1$, the solution $\varphi_{1}(\xi)$
remains valid  at $\mu\xi^{1/2}\gg 1$, as long as $\xi \ll 1$. It has,
therefore, a common region of validity with the WKB solutions. We use the
asymptotic expansion \cite{bateman}
\begin{equation}\label{bessel}
J_{1}(z)\simeq\sqrt{\frac{2}{\pi{z}}}\,\cos\left(\frac{3\pi}{4}-z\right)\;\;\;\mbox{at}\;\;\;z\gg{1}
\end{equation}
and obtain, up to a constant factor,
\begin{eqnarray}
\varphi_{1}(\xi) &\simeq& \xi^{1/4}\sin\left(\mu \sqrt{2 \xi} -\frac{\pi}{4}\right) \nonumber\\
&=&(1-x)^{1/4}\sin\left[\mu\sqrt{2(1-x)}-\frac{\pi}{4}\right]\,. \label{asymp}
\end{eqnarray}
Now we expand the even WKB solution (\ref{wkbsolann}) at $1-x\ll{1}$:
\begin{equation}\label{even_expand}
 \varphi_{ev}(x)  \simeq (1 - x)^{1/4}\sin\left[\mu\sqrt{2(1 - x)} - \frac{\mu\pi}{2}
- \frac{\pi}{2}\right]\,.
\end{equation}
Matching the asymptotes (\ref{asymp}) and (\ref{even_expand}), we obtain the
discrete spectrum eigenvalues corresponding to the even eigenfunctions:
\begin{equation}\label{eigen_even}
\mu=2k-\frac{1}{2}\,,
\end{equation}
where $k \gg 1$ is an integer. In a similar way we obtain the discrete spectrum
for the odd eigenfunctions:
\begin{equation}\label{eigen_odd}
\mu=2k+\frac{1}{2}\,.
\end{equation}
The eigenvalues (\ref{eigen_even}) and (\ref{eigen_odd}) coincide, in the
leading and subleading orders in $k\gg{1}$, with the exact eigenvalues
$(4k^2-2k)^{1/2}$ and $(4k^2+2k)^{1/2}$, respectively, see Appendix.
Furthermore, the corresponding WKB eigenfunctions provide an accurate
approximation, see Fig. \ref{eigen}, to the exact eigenfunctions [which are
orthogonal, with respect to the inner product with the weight function
$w(x)=(1-x^2)^{-1}$, on the interval $|x|<1$, and form a complete set]. We
normalize the approximate eigenfunctions $u_{l}(x)$ by demanding
\begin{equation}\label{norm}
   \int_{-1}^{1} u_{l}^2(x)
w(x)\,dx=1\,.
\end{equation}
The normalized even WKB eigenfunctions are
\begin{equation}\label{wkbeigenfunction}
u_{2k}(x)=\left\{\begin{array}{ll}
u_{2k}^{(a)}(x)&=(-1)^{k}\sqrt{\frac{2}{\pi}}(1-x^{2})^{1/4}\nonumber\\
&\times \cos\left[\mu_{k}\arcsin x\right]\nonumber\\
&\;\;\;\;\;\;\;\mbox{for}\;
0 \leq |x|<1-\varepsilon\,,\nonumber\\
u_{2k}^{(b)}(x)&= - \sqrt{2\mu_{k} (1-|x|)} \nonumber\\
&\times J_{1} \left[ \mu_{k} \sqrt{2(1 - |x|)}\right]\nonumber\\
&\;\;\;\;\;\;\;\mbox{for} \; 1-\varepsilon<|x|\leq{1}\,,\end{array}\right.
\end{equation}
where $1/k^{2}\ll\varepsilon\ll{1}$, $\mu_{k}=2k-1/2$, and $k\gg{1}$. For
$1/\mu_k^2\ll 1-|x| \ll {1}$ the function $u_{2k}^{(a)}(x)$ coincides, in the
leading order, with $u_{2k}^{(b)}(x)$. It is sufficient to use only the
$u_{2k}^a(x)$ asymptotes in the normalization integral (\ref{norm}).

\section{Initial value problem and calculation of the probabilities}

Putting everything together, we can write the WKB solution of the initial value
problem for Eq. (\ref{anndiffeq}) as
\begin{equation}\label{wkbgeneral}
G(x,t)\simeq G_s(x)+\sum_{l>0} a_l \exp\left(-\frac{\mu_l^2 \lambda
t}{2}\right)\,u_l(x)\,,
\end{equation}
where each constant $a_l$ is equal to the inner product [with the weight
function $w(x)=(1-x^2)^{-1}$] of the respective normalized eigenfunction
$u_l(x)$ and the function $G_0 (x)-G_s(x)$. One can see that the populations of
the eigenstates with $l>0$ of this non-Hermitian ``quantum mechanics" are
decaying exponentially in time.

Assume, for concreteness, that the initial number of particles is fixed and
equal to $n_{0}=2k_{0}\gg{1}$, where $k_{0}$ is integer. In this case [see Eq.
(\ref{incond})] $G_{0}(x)=x^{2k_{0}}$, and one only needs the even
eigenfunctions (\ref{wkbeigenfunction}). Now we can use the orthonormality
relation (\ref{norm}) and compute the coefficients $a_k$. After a lengthy
algebra we obtain
\begin{equation}\label{wkbcoefffinal}
a_{k}\simeq\frac{2}{\sqrt{\mu_{k}}}\,e^{-\frac{k(2k-1)}{2k_{0}}}\,,
\end{equation}
where we have assumed $k\lesssim\sqrt{k_0}\ll k_0$, as justified below. In the
leading order in $1/k\ll 1$ Eq. (\ref{wkbcoefffinal}) coincides with the
corresponding asymptotics (\ref{aapprox}) of the exact result. Now Eq.
(\ref{wkbgeneral}) becomes
\begin{equation}\label{e19}
G(x,t)\simeq 1+\sum_{k=1}^{\infty}\frac{2}{\sqrt{\mu_{k}}}
\,e^{\frac{-k(2k-1)}{\bar{N}(t)}}u_{2k}(x)\,,
\end{equation}
where
$$
\bar{N}(t)\equiv\frac{2k_0}{1+2k_0\lambda t}
$$
is the average number of particles at time $t$ according to the mean-field
theory. Though the sum in Eq. (\ref{e19}) formally runs to infinity, the
dominant contribution comes from terms with $k\lesssim\sqrt{k_{0}}\ll k_{0}$,
while the rest of terms give only exponentially small corrections.

To recover $P_n(t)$, for $1\ll n \lesssim \sqrt{k_0}$, we use Eq.
(\ref{probformula}):
\begin{equation}\label{probwkbformula}
P_{n}(t) \simeq \frac{1}{n!}
\sum_{k=1}^{\infty}\frac{2}{\sqrt{\mu_{k}}}\left.\,e^{\frac{-k(2k-1)}{\bar{N}(t)}}\;
\frac{\partial^{n}u_{2k}(x)}{\partial\, x^{n}}\right|_{x=0}.
\end{equation}
As our WKB approximation assumes $\mu_k =2 k-1/2 \gg 1$, the first few terms of
the sum in Eq. (\ref{probwkbformula}) may seem inaccurate. It turns out,
however, that the sum actually starts from $k=n/2$, see below. Therefore, at
$1\ll n \lesssim \sqrt{k_0}$ all the terms of the sum in Eq.
(\ref{probwkbformula}) are accurate.

Equation (\ref{probwkbformula}) is the central result of the stationary WKB
approximation for the binary annihilation problem. To compute the $n$-th
derivative of the WKB eigenfunctions $u_{2k}(x)$, entering Eq.
(\ref{probwkbformula}), we can analytically continue $u_{2k}(x)$ into the
complex plane. By virtue of the Cauchy theorem
\begin{equation}\label{complexint}
I \equiv \frac{1}{n!}\left.\frac{\partial^{n}u_{2k}(x)}{\partial\,
x^{n}}\right|_{x=0}=\frac{1}{2\pi i}\oint_C \frac{u_{2k}(z)dz}{z^{n+1}}\,,
\end{equation}
where the integration is performed over a closed contour $C$ in the complex
$z$-plane around the pole $x=0$ inside the region of analyticity of $u_{2k}(x)$.
We can put here $u_{2k}(z)=u_{2k}^{(a)}(z)$, since the main contribution to the
integral, as shown below, comes from the region far from the points $x=\pm 1$,
in the vicinity of which $u_{2k}^{(a)}(z)$ is inaccurate. We obtain
\begin{equation}
I=\frac{(-1)^{k}}{\sqrt{2}\pi^{3/2} i} \,\mbox{Im} \left[
\oint_{C}g(z)\,e^{(n-1)f(z)} dz\right]\,, \label{complexint1}
\end{equation}
where $f(z)=iA\arcsin(z)-\ln(z)$, $g(z)=\frac{(1 - z^{2})^{1/4}}{z^{2}}$ and
$A=\mu_{k}/(n-1)$. As $n \gg 1$, the integral in (\ref{complexint1}) can be
evaluated using the saddle point approximation \cite{orszag}. This is done by
deforming the contour $C$, so that it passes through the saddle point
$z_*=x_*+iy_*$, where $u(x,y)=\mbox{Re} [f(z)]$ obtains its maximum, and
consequently $v(x,y)=\mbox{Im} [f(z)]$ is constant. By choosing the contour at
the saddle point to be parallel to the direction of the steepest descent of
$u(x,y)$, we can replace the integration over the complex plane by integration
over the real axis, having to multiply the result by a constant phase: the value
of $v(x,y)$ at the saddle point. The saddle point can be found from the equation
$f^{\prime}(z)=0$. For $k<n/2$ (that is, $A<1$), the saddle point lies on the
real axis: $z_{*}^{(1)}=[1-A^{2}]^{-1/2}$, whereas for $k\geq{n/2}$ (that is,
$A>1$) it lies on the imaginary axis: $z_{*}^{(2)}=-i[A^{2}-1]^{-1/2}$. In each
of these cases Eq. (\ref{complexint1}) becomes \cite{orszag}
\begin{equation}\label{complexsol}
I=\frac{(-1)^{k}}{\sqrt{2}\pi^{3/2} i}
\,\mbox{Im}\left[\frac{\sqrt{2\pi}g(z_{*})e^{(n-1)f(z_{*})}e^{i\alpha}}{\sqrt{(n-1)|f''(z_{*})|}}\right]\,,
\end{equation}
where $\alpha$ is the angle of the contour with respect to the positive real
axis at the saddle point, where the contour is chosen to be parallel to $ -
\nabla u(x,y)$.

\begin{figure}
\includegraphics[width=8cm,clip=]{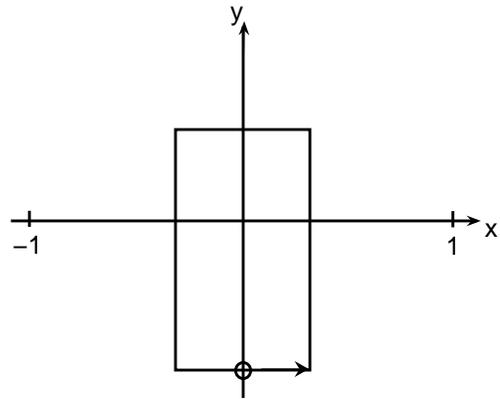}
\caption{A possible contour $C$, see Eq. (\ref{complexint1}). The
direction of the contour in the vicinity of the saddle point
$z=z_{*}^{(2)}$ (marked by the circle) is chosen to be along $-
\nabla u(x,y)$ at $z_{*}^{(2)}$, implying in this case $\alpha=0$.
As $n\gg{1}$, the contribution of the rest of the contour (as long
as it encircles $z=0$) is exponentially small.} \label{contour}
\end{figure}

This procedure yields markedly different results in the cases of $A<1$ ($k<n/2$)
and $A>1$ ($k\geq n/2$). For $A<1$, upon substituting $z_{*}^{(1)}$ and
deforming the contour so that $\alpha=\pi/2$ in its vicinity, we realize that
the result inside the brackets in Eq. (\ref{complexsol}) is real which yields
$I=0$. Therefore, the saddle-point asymptote Eq. (\ref{complexsol}) predicts
that the $n$-th derivative of the eigenfunctions $u_{2k}(x)$ vanishes for
$k<n/2$, so the sum in Eq. (\ref{probwkbformula}) starts from $k=n/2$. This
could be expected, as the same kind of behavior is exhibited by the
\textit{exact} eigenfunctions $q_{2k}(x)$, which are polynomials of order $2k$,
see Appendix.

\begin{figure}[h]
\includegraphics[width=8.5cm,clip=]{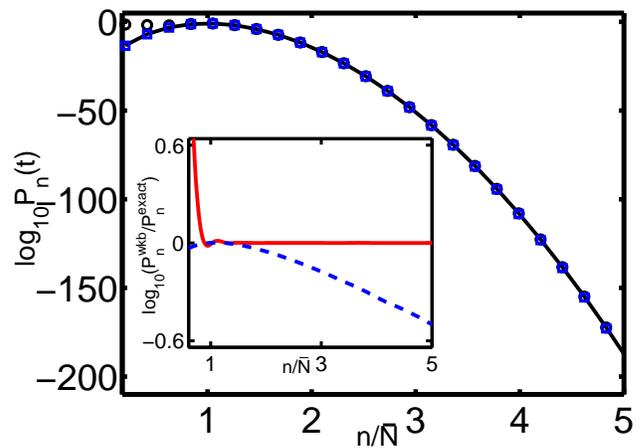}
\caption{(Color online) The decimal logarithm of $P_{n}(t)$ as a
function of $n/\bar{N}$ for $\lambda{t}=0.02$ and $n_{0}=10^3$.
Shown are the exact probabilities (\ref{probeven}) (the solid line),
the stationary WKB probabilities (\ref{probwkbformula}) (the empty
circles) and the time-dependent WKB probabilities (\ref{Pkam}) (the
blue squares). Inset shows the logarithm of the ratio of the
stationary WKB probabilities and the exact ones (the red solid
line), and the logarithm of the ratio of the time-dependent WKB
probabilities and the exact ones (the blue dashed line), versus
$n/\bar{N}$. As $n/\bar{N}$ grows, the time-dependent WKB solution
deteriorates, whereas when $n$ goes down below $\bar{N}$, the
stationary WKB solution deteriorates.} \label{statwkbexact50}
\end{figure}

\begin{figure}[h]
\includegraphics[width=8.5cm,clip=]{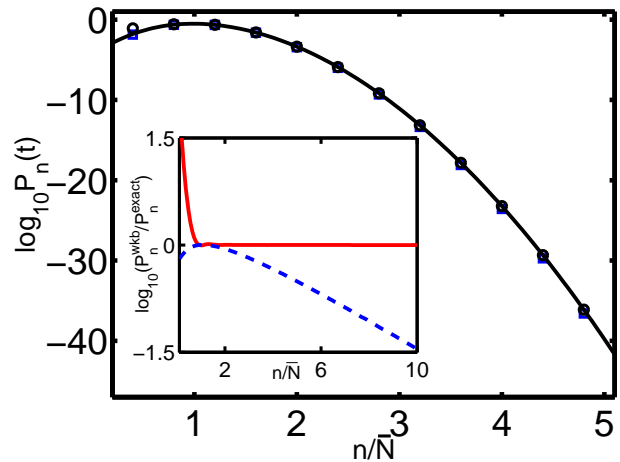}
\caption{Same as in Fig. \ref{statwkbexact50}, but for
$\lambda{t}=0.1$.} \label{statwkbexact10}
\end{figure}

\begin{figure}[h]
\includegraphics[width=8.5cm,clip=]{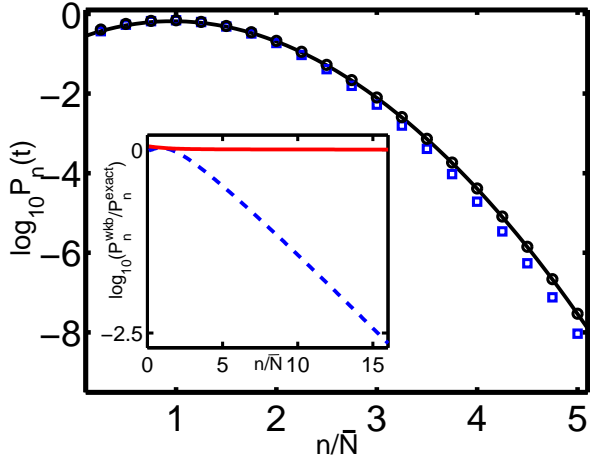}
\caption{Same as in Fig. \ref{statwkbexact50}, but for
$\lambda{t}=0.5$. One can clearly see that the time-dependent WKB
result is already inaccurate at $n \gtrsim \bar{N}$, while the
stationary WKB result keeps its high accuracy there.}
\label{statwkbexact2}
\end{figure}

\begin{figure}[h]
\includegraphics[width=8cm,clip=]{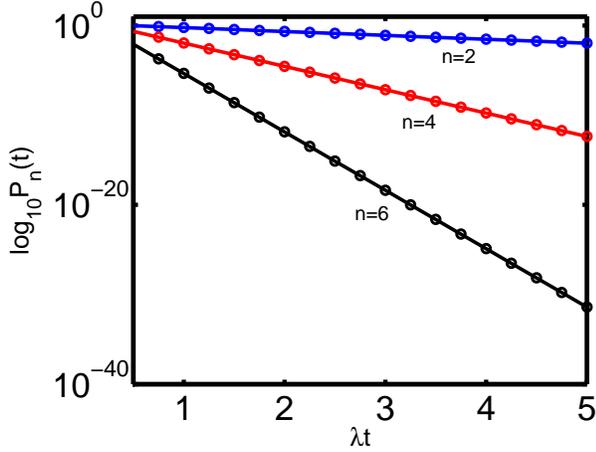}
\caption{Long time asymptotics of $P_{n}(t)$. Shown is the decimal logarithm of
$P_{n}(t)$ as a function of $\lambda t$ for $n_{0}=10^3$ and $n=2,4,$ and $6$.
The circles denote the stationary WKB solution [Eq. (\ref{probwkblongtimes})],
the solid lines denote the exact solution [Eq. (\ref{probeven})]. Notice
excellent agreement even for small $n$.}\label{longtimeasym}
\end{figure}

After some algebra, Eq. (\ref{complexsol}) yields, for $k \geq n/2$
\begin{equation}
\label{derivative} I \simeq \frac{(-1)^{k-n/2}e^{1/4} k^{3/2}2^{n}
\left(k+\frac{n}{2}\right)^{k+\frac{n}{2}-1}}{\pi\,n^{n
+\frac{1}{2}}\,\left(k-\frac{n}{2}+\frac{1}{4}\right)^{k-\frac{n}{2}+\frac{1}{2}}}\,,
\end{equation}
where we have substituted $z_{*}^{(2)}$ for the saddle point. A possible contour
$C$ is shown in Fig. \ref{contour}. We can see now that, as the saddle point
$z_{*}$ lies on the imaginary axis, the contour does not have to come close to
$x=\pm{1}$, thus justifying the use of $u_{2k}^{(a)}(z)$ for $u_{2k}(z)$. The
saddle point approximation is only valid when $|f'''(z_*)|/|f''(z_*)|^{3/2}\ll
\sqrt{n}$ \cite{orszag}. For $k\geq{n/2}$ this requirement is equivalent to
$k-n/2\gg 1$. We will see shortly, however, that the results remain quite
accurate even for $k-n/2={\cal O}(1)$. Now we can rewrite Eq.
(\ref{probwkbformula}) as
\begin{eqnarray}
P_{n}(t)&\simeq& \frac{e^{1/4} 2^{n+\frac{1}{2}}}{\pi\,n^{n+\frac{1}{2}}}
\sum_{k=\frac{n}{2}}^{\infty}\frac{(-1)^{k-\frac{n}{2}}\,k\left(k+\frac{n}{2}\right)^{k
+\frac{n}{2}-1}}{\left(k-\frac{n}{2}+\frac{1}{4}\right)^{k-\frac{n}{2}+\frac{1}{2}}}
\nonumber\\
&\times&\,e^{\frac{-k(2k-1)}{\bar{N}(t)}}. \label{probwkbfinal}
\end{eqnarray}
We immediately notice that this formula is very similar to Eq.
(\ref{probevenapprox1}). Moreover, the two expressions \textit{coincide} if we
rewrite the factor $(k-n/2)^{k-n/2+1/2}$ in the denominator of Eq.
(\ref{probevenapprox1}) as $(k-n/2+1/4-1/4)^{k-n/2+1/2}$, assume $k-n/2\gg 1$
and use the asymptote $(1+\xi/u)^u \simeq e^{\xi}$ at $u\gg 1$. As Eq.
(\ref{probevenapprox1})  gives an accurate approximation to the exact
probabilities at $n \sim \bar{N}$ (see Appendix), the same is true for the
stationary WKB result (\ref{probevenapprox1}). For $n\gg \bar{N}$ it suffices to
take into account only the first term, $k=n/2$ in the sum of Eq.
(\ref{probwkbfinal}) which yields
\begin{equation}\label{probwkblongtimes}
P_{n}(t)\simeq\frac{2^{n+\frac{1}{2}}\,e^{\frac{1}{4}}}{\pi\sqrt{n}}\,e^{\frac{-n(n-1)}{2\bar{N}(t)}}\,.
\end{equation}
This asymptote coincides, up to a factor $\sqrt{2/\pi}\,e^{1/4}\simeq{0.976}$,
with Eq. (\ref{probevenapproxlong}) that gives an accurate approximation to the
exact probabilities in this limit, see Appendix. Therefore, in the region of
$n\gtrsim\bar{N}$ the stationary WKB theory is accurate, as can be seen from
Figures \ref{statwkbexact50}-\ref{statwkbexact2}. Importantly, the asymptote
(\ref{probwkblongtimes}) does not demand $\bar{N}\gg 1$ and therefore remains
valid at long times, $\lambda t\gtrsim 1$, when the average number of particles
is already small, see Fig. \ref{longtimeasym}.

\section{Time-dependent semiclassical solution versus exact solution}

The time-dependent semiclassical approximation, suggested by Elgart and Kamenev
\cite{kamenev}, differs from the stationary WKB approximation in that it deals
semiclassically with the original non-Hermitian Schr\"{o}dinger equation
(\ref{generaleq}) (a partial differential equation), rather than with the set of
ordinary differential equations obtained by the separation of variables in Eq.
(\ref{generaleq}). Using the ansatz $G(x,t)=\exp[-S(x,t)]$ and neglecting the
$\partial_{xx} S$ term, one arrives at a Hamilton-Jacobi equation. For the
binary annihilation problem this equation is \cite{kamenev}
\begin{equation}\label{HJ}
    \frac{\partial{S}}{\partial{t}} =
\frac{\lambda}{2}(x^{2}-1)\left(\frac{\partial{S}}{\partial{x}}\right)^{2}\,.
\end{equation}
Elgart and Kamenev \cite{kamenev} found an exact solution to this equation:
$$
S(x,t)= \frac{1}{2} \,\bar{N}(t)\arccos^2 x\,.
$$
The respective generating function
\begin{equation}\label{elgart0}
G(x,t)=\exp\left[-\frac{1}{2} \,\bar{N}(t)\arccos^2 x\right]
\end{equation}
obeys the boundary condition  (\ref{universalBC}) exactly, and the initial
condition $G(x,t=0) = x^{n_0}$ with a high accuracy as long as $n_0 \gg 1$. The
probabilities $P_{n}(t)$ can be calculated from the equation
\begin{equation}\label{kam}
\left.P_{n}(t)=\frac{1}{n!}\frac{\partial^{n}G(x,t)}{\partial
x^{n}}\right|_{x=0}\,= \frac{1}{2\pi{i}}\oint\frac{dz}{z^{1+n}}G(z,t)\,,
\end{equation}
where the integration is performed over a closed contour including $z=0$ in the
complex $z$ plane, inside the region of analyticity of $G(z,t)$. For $n\gg{1}$
and $\bar{N}(t)\gg 1$, the integral can be evaluated by the saddle point
approximation, and the result is
\begin{equation}\label{Pkam}
P_n(t)\simeq \sqrt{\frac{1-x_s^{2}}{2\pi\left[n-\bar{N}(t)x_s^{2}\right]}}\, e^{
- \bar{N}(t) \left(\frac{1}{2} \arccos^2 x_s+\frac{n}{\bar{N}(t)} \ln
x_s\right)},
\end{equation}
where $x_s=x_s(n/\bar{N})$ is the root of the saddle-point equation $x_s
(1-x_s^2)^{-1/2} \arccos x_s=n/\bar{N}$ \cite{kamenev}. For $n\ll\bar{N}(t)$ one
obtains $x_s\simeq 2n/(\pi\bar{N})$, so
\begin{equation}\label{asymp1}
\ln P_{n}(t)\simeq n\ln\frac{\pi\bar{N}(t)}{2n}-\frac{\pi^{2}\bar{N}(t)}{8}+n
\end{equation}
(in the corresponding asymptote of Ref. \cite{kamenev} the term $n$ is missing.)

For $n\simeq\bar{N}(t)$, $x_s\simeq 1-(3/2)(1-n/\bar{N})$, so
\begin{equation}\label{asymp2}
\ln P_{n}(t)\simeq -\frac{3[n-\bar{N}(t)]^{2}}{4\bar{N}(t)}\,.
\end{equation}

Finally, for $n\gg \bar{N}(t)$, $x_s\simeq (1/2) \,e^{n/\bar{N}}$,
so
\begin{equation}\label{asymp3}
\ln P_{n}(t)\simeq n\ln 2-\frac{n^{2}}{2\bar{N}(t)}\,.
\end{equation}

\begin{figure}[h]
\includegraphics[width=8cm,clip=]{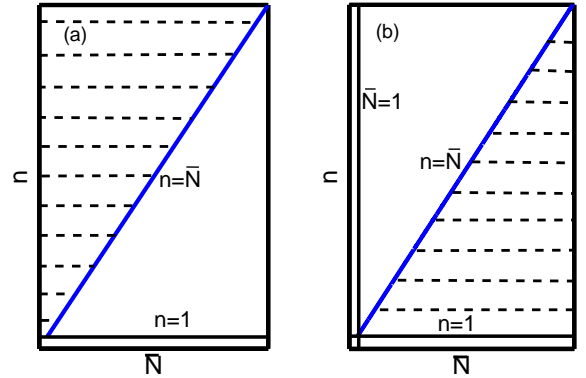}
\caption{The dashed regions schematically show the validity of each
of the two semiclassical theories on the parameter plane $(\bar{N},
n)$. Figure a shows the validity domain $n\gtrsim\bar{N}$ and $n\gg
1$ of the stationary WKB theory. An additional condition for its
validity  is $n \lesssim \sqrt{k_0}\ll k_0$ (not shown). Figure b
shows the validity domain $1\ll n\lesssim\bar{N}$ of the
time-dependent WKB theory.} \label{phasespace}
\end{figure}

What are the applicability conditions of the time-dependent semiclassical
approximation? The above calculations required $n_0=2 k_0\gg 1$, $n\gg 1$ and
$\bar{N}\gg 1$. To find out whether there is an additional condition, let us
consider the $n\gg\bar{N}(t)$ asymptote of the \textit{exact} result [Eq.
(\ref{probevenapproxlong})]:
\begin{equation}\label{logexactprob}
\ln P_{n}(t)\simeq n\ln 2-\frac{n^{2}}{2\bar{N}(t)}+\frac{n}{2\bar{N}(t)}\,.
\end{equation}
A comparison of Eqs. (\ref{asymp3}) and (\ref{logexactprob}) shows that the
time-dependent WKB probability $P_n(t)$ lacks a large term $n/[2\bar{N}(t)]$ in
the exponent, and therefore it greatly underestimates rare events with $n\gg
\bar{N}(t)$. This effect can been seen in Figs.
\ref{statwkbexact50}-\ref{statwkbexact2}. On the other hand, in the region
$n\lesssim\bar{N}(t)$, the time-dependent WKB theory yields a good approximation
to the exact result, as it circumvents the summation of large terms of
alternating sign.

Therefore, based on the analytical and numerical comparisons, we conclude that
the time-dependent semiclassical approximation is accurate at $1\ll n \lesssim
\bar{N}$. This obviously implies $\bar{N} \gg 1$, that is not too long times:
$\lambda t \ll 1$.  The stationary WKB approximation is accurate for $n\gtrsim
\bar{N}$ for \textit{any} $\bar{N}$, that is for all times. These results are
illustrated in Figs. \ref{statwkbexact50}-\ref{statwkbexact2} which show
$P_n(t)$ versus $n/\bar{N}$: the exact result (\ref{probeven})  and the
predictions of each of the two approximations, Eq. (\ref{probwkbformula}) and
Eq. (\ref{Pkam}). The validity domains of each of the two semiclassical
approximations in the parameter plane ($\bar{N},n$) are shown in Fig.
\ref{phasespace}.

\section{Summary and Discussion}

We developed a spectral formulation and a stationary WKB approximation for
calculating the probabilities of rare events in systems of reacting particles
with infinite-range interaction which are describable by a master equation. We
extended the exact analytical solution of the binary annihilation problem $2A
\to \emptyset$ and used is as a benchmark for testing the stationary WKB
approximation and a recent time-dependent WKB approximation due to Elgart and
Kamenev \cite{kamenev}. In theory the stationary WKB approximation is always
more accurate than a time-dependent WKB approximation. In practice this
advantage is indeed realized in the regimes where the superposition of the
different ``quantum states" of the system is dominated by a small number of
terms. On the contrary, when many ``quantum states" are involved, virtually
\textit{any} approximation to the ``wave functions" of individual states may
alter the precise destructive interference between the different states and
cause large errors. In such cases the time-dependent WKB approximation
\cite{kamenev}, which effectively sums over the ``quantum states" without
dealing with them explicitly, can be advantageous.

WKB approximation alone is insufficient for calculating the life-time
probability distribution of systems which exhibit extinction. This quantity is
encoded in $P_0(t)\equiv G(x=0,t)$ which involves a sum over \textit{all}
``quantum states", including the lowest ones (see the final part of Appendix).
Still, the spectral formulation can be very useful here: it clearly identifies
the lowest  states and provides a proper framework for calculating their
eigenvalues and eigenfunctions: for example, by a variational method.

This work dealt with the case when the operator $\hat{L}$ is of the second
order, and the standard machinery of Sturm-Liouville theory is therefore
available. The spectral formulation itself, however, is equally applicable to
higher-order operators. Furthermore, it is well known that ``WKB analysis is not
sensitive to the order of a differential equation"  (Ref. \cite{orszag}, p. 496)
which paves the way to generalizations of the theory to more complicated
reaction kinetics.

Finally, we have restricted ourselves in this paper to a single species. The
generating function formalism, however, is applicable to any number of species
[see Eqs. (\ref{genfunmany})-(\ref{universalmanyBC})]. Already for two species
some qualitative changes are possible. Indeed, for two species, the underlying
classical phase space, described by the Hamiltonian of the problem, is
four-dimensional. If energy is the only integral of motion, the classical motion
is non-separable and, in general, chaotic \cite{gutzwiller}. Therefore, for
highly-excited states, where classical mechanics and stationary WKB
approximation are relevant, the spectral formulation  may bring about an
extension of quantum chaos to these non-Hermitian ``quantum" systems.

\begin{acknowledgments}
We are very grateful to Alex Kamenev,  Vlad Elgart, and Len Sander for helpful
discussions. B.M. is grateful to the Michigan Center for Theoretical Physics,
University of Michigan, where this work started. The work was supported by the
Israel Science Foundation (grant No. 107/05).
\end{acknowledgments}

\section*{Appendix}
\renewcommand{\theequation}{A\arabic{equation}}
\setcounter{equation}{0} Here we briefly review and extend the exact solution of
the binary annihilation problem, obtained by McQuarrie \textit{et al.}
\cite{McQuarrie}. Let the initial state correspond to a fixed and even number of
particles, so one only needs the even eigenfunctions of Eq. (\ref{oanndiffeq}),
with the boundary conditions $\phi(x=\pm{1})=0$. The exact even eigenfunctions
are \cite{Abramowitz}
$$
\phi_{2k}(x)={\cal P}_{2k}(x) - {\cal P}_{2k - 2}(x)\,,\;\;\;\;k=1,2,\dots\,,
$$
where ${\cal P}_{l}(x)$ is the Legendre polynomial of order $l$. The
corresponding exact eigenvalues are $\mu=(4k^2-2k)^{1/2}$. The
\textit{normalized} even eigenfunctions, see Eq. (\ref{norm}), are
$$
q_{2k}(x)=\sqrt{\frac{k(2k-1)}{4k-1}}\left[{\cal P}_{2k}(x)-{\cal
P}_{2k-2}(x)\right]\,.
$$
The exact solution of an initial value problem for $G(x,t)$ can be written as
\begin{equation}\label{gensoleven}
G(x,t)=1+\sum_{k=1}^{\infty}A_{k}q_{2k}(x)e^{-k(2k - 1)\lambda t}\,,
\end{equation}
where the coefficients $A_{k}$ are determined by the initial data $G_0(x)$. For
the initial data we are interested in $n_{0}=2k_{0}$, where $k_{0}=0,1,2,...$,
therefore, $G_0(x)=x^{2k_{0}}$. Using the orthogonality relations of the
Legendre polynomials, we obtain, after some algebra,
\begin{equation}\label{aexact}
A_{k}=\sqrt{\frac{4k-1}{k(2k-1)}}\;\frac{\Gamma(1+k_{0})
\Gamma\left(\frac{1}{2}+k_{0}\right)}{\Gamma\left(\frac{1}{2}+k_{0}+k\right)\Gamma(k_{0}-k+1)}\,,
\end{equation}
where $\Gamma(\cdot\cdot\cdot)$ is the gamma function. Owing to the presence of
factor $\Gamma(k_{0}-k+1)$ in the denominator, $A_{k}$ vanishes for $k>k_{0}$.
Therefore, the sum in Eq. (\ref{gensoleven}) is finite in this case, and it ends
at $k=k_0$. The average number of particles at time $t$, $\bar{n}(t)$, can be
found from the following relation:
$$
\bar{n}(t)=\sum_{n=0}^{\infty}nP_{n}(t)=\left.\frac{\partial G(x,t)}{\partial
x}\right|_{x=1}\,.
$$
Using Eq. (\ref{gensoleven}), we obtain
\begin{equation}\label{averageeven}
\bar{n}(t)=\sum_{k=1}^{k_{0}}\sqrt{k(2k-1)(4k-1)}\,A_{k} \,e^{-k(2k-1)\lambda
t}\,.
\end{equation}
Equations (\ref{gensoleven})-(\ref{averageeven}) coincide, up to notation, with
the results of McQuarrie \textit{et. al.} \cite{McQuarrie} [they also calculated
the second moment $\bar{n^2} (t)$]. We now extend the exact theory in three
directions. First, we obtain some useful approximations for a large number of
particles. Second, we calculate the probabilities $P_n(t)$ and their approximate
asymptotics in different regimes. We use these approximations while comparing
the exact solution (i) with our stationary WKB results, and (ii) with the
time-dependent semiclassical approximation of Elgart and Kamenev \cite{kamenev}.
Third, we find the probability distribution of lifetimes of the particles in
this system, and the average extinction time.

When $k \lesssim \sqrt{k_{0}}$ (see below), we can reduce Eq. (\ref{aexact}) to
\begin{equation}\label{aapprox}
A_{k}\simeq\sqrt{\frac{4k-1}{k(2k-1)}}\,e^{-\frac{k}{2k_{0}}(2k-1)}\,
\end{equation}
which  yields the following approximation for $G(x,t)$:
\begin{equation}\label{gensolevenapprox}
G(x,t)\simeq{1}+\sum_{k=1}^{k_{0}}\sqrt{\frac{4k-1}{k(2k-1)}}\,q_{2k}(x)\,e^{-\frac{k(2k
- 1)}{\bar{N}(t)}}\,.
\end{equation}
Here
$$
\bar{N}(t) \equiv\frac{2k_{0}}{1+2k_{0}\lambda t}\,,
$$
which is the mean-field result for the average number of particles. Equation
(\ref{gensolevenapprox}) is valid for \textit{all} times, as the dominant
contribution to the sum comes from terms with $k\lesssim\sqrt{k_{0}}\ll k_{0}$,
while the rest of terms give only exponentially small corrections. The average
number of particles (\ref{averageeven}) can be approximated as
\begin{equation}\label{avgapproxeven}
\bar{n}(t)\simeq \sum_{k=1}^{k_{0}}(4k-1)\,e^{-\frac{k(2k - 1)}{\bar{N}(t)}}\,.
\end{equation}
For short times, $\lambda t \ll 1$,  $\bar{N}(t)\gg 1$, so the summation in Eq.
(\ref{avgapproxeven}) can be replaced by integration. Moving the upper limit  to
infinity (which only causes an exponentially small error), we obtain
\begin{equation}
\label{avgapproxeven1}
  \nonumber \bar{n}(t) \simeq \int_{1}^{\infty}(4k-1)\,e^{-\frac{k(2k -
1)}{\bar{N}(t)}}\,dk \simeq\bar{N}(t)\,,
\end{equation}
the mean-field result. In the long-time limit  $\lambda t \gg 1$,
$\bar{N}(t)\simeq (\lambda t)^{-1} \ll 1$, the term $k=1$ in Eq.
(\ref{avgapproxeven}) is dominant, and we obtain $\bar{n}(t)\simeq 3\,
e^{-\lambda t}$.

Now we employ Eq. (\ref{probformula}) of the main part of the paper to calculate
the probabilities $P_{n}(t)$. After some algebra we obtain the exact result:
\begin{equation}\label{probeven}
P_{n}(t)=\delta_{0n} + \frac{2^{n - 1}}{n!}\sum_{k=r}^{k_{0}} C_{e}(k,n)e^{ -
k(2k - 1) \lambda t}\,,
\end{equation}
where $n$ is assumed to be even, $r=\mbox{max}(1,n/2)$, and
\begin{equation}\label{probcoeffeven}
C_{e}(k,n) = {A_{k}}\left[\frac{( - 1 )^{k-\frac{n}{2}}(4k - 1)\Gamma[k -
\frac{1}{2} + \frac{n}{2}]}{\sqrt{\pi}\left(k - \frac{n}{2}\right)! }\right]\,.
\end{equation}
At $n>0$ the sum in Eq. (\ref{probeven}) starts from $k=n/2$. This is because
the eigenfunctions $q_{2k}(x)$ are polynomials of order $2 k$, so the $n$-th
derivative of $q_{2k}(x)$ vanish at $n>2k$.

Going to the limit of $n \gg 1$ and $k \lesssim \sqrt{k_0}$, and using the
approximation (\ref{aapprox}) for $A_k$, we can rewrite $C_{e}(k,n)$ as
\begin{equation}\label{probcoeffapprox}
C_{e}(k,n)\simeq\frac{( - 1 )^{k -
\frac{n}{2}}2^{5/2}k\left(k+\frac{n}{2}\right)^{k + \frac{n}{2} - 1}}{\left(k -
\frac{n}{2}\right)! \,e^{k+\frac{n}{2}}}\,e^{-\frac{k(2k - 1)}{2k_{0}}}\,.
\end{equation}
Therefore, at $1\ll n \lesssim \sqrt{k_0}$, $P_{n}(t)$ becomes
\begin{eqnarray}
P_{n} (t) \simeq \frac{2^{n + 1}}{n^{n + \frac{1}{2}}}\sum_{k =
\frac{n}{2}}^{k_{0}} \frac{( - 1 )^{k - \frac{n}{2}}k\left(k +
\frac{n}{2}\right)^{k + \frac{n}{2} - 1}}{\sqrt{\pi}\left(k -
\frac{n}{2}\right)! \,e^{k -
\frac{n}{2}}}\,e^{\frac{-k(2k-1)}{\bar{N}(t)}}\,.\nonumber\\
\label{probevenapprox}
\end{eqnarray}
This expression can be simplified drastically for $n\gg\bar{N}(t)$. Here the sum
can be accurately approximated by its first term $k=n/2$, and we obtain
\begin{equation}\label{probevenapproxlong}
P_{n}(t)\simeq\frac{2^{n}}{\sqrt{\pi{n}}}\,e^{\frac{-n(n-1)}{2\bar{N}(t)}}\,.
\end{equation}
This asymptote coincides, in the limit of $1\ll n \lesssim \sqrt{k_0}$, with the
first term of the sum in the exact result (\ref{probeven}).

\begin{figure}[h]
\includegraphics[width=8cm,clip=]{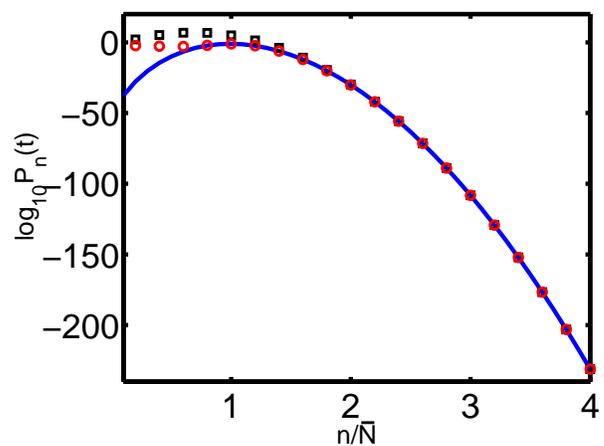}
\caption{The exact probabilities (\ref{probeven}) (the solid line),
the approximate probabilities (\ref{probevenapprox}) (the circles),
and their asymptotics (\ref{probevenapprox1}) and
(\ref{probevenapproxlong}) (the squares), for $n_{0}=10^{4}$ and
$\bar{N}(t=0.01)\simeq 99.5$. The agreement is good for
$n\gtrsim\bar{N}(t)$.} \label{figappend}
\end{figure}

In the cases of $n\ll \bar{N}(t)$ and $n \sim \bar{N}(t)$ the leading
contribution to the sum in Eq. (\ref{probevenapprox}) comes from terms for which
$k-n/2\gg 1$ which makes it possible to use the Stirling formula for the factor
$\left(k - n/2\right)!\,$ and arrive at
\begin{eqnarray}
P_{n} (t) \simeq\frac{2^{n + \frac{1}{2}}}{n^{n + \frac{1}{2}}}\sum_{k >
\frac{n}{2}}^{k_{0}} \frac{( - 1 )^{k - \frac{n}{2}}k\left(k +
\frac{n}{2}\right)^{k + \frac{n}{2} - 1}}{\pi\left(k - \frac{n}{2}\right)^{k -
\frac{n}{2} + \frac{1}{2}}}
\,e^{\frac{-k(2k-1)}{\bar{N}(t)}}.\nonumber\\
\label{probevenapprox1}
\end{eqnarray}

Now let us go back to Eq. (\ref{probevenapprox}). According to our analysis, it
should be accurate for $1\ll n  \lesssim \sqrt{k_0}$. A numerical comparison
with the exact result (\ref{probeven}) [see Fig. \ref{figappend}] shows,
however, that Eq. (\ref{probevenapprox}) is accurate only at $\bar{N}\lesssim n
\lesssim \sqrt{k_0}$. At $n\lesssim \bar{N}$ the agreement rapidly deteriorates.
The disagreement stems from the fact that the sum in Eq. (\ref{probeven})
consists of terms of alternating sign. In the region of $n \lesssim \bar{N}$,
$P_n(t)$ is much smaller than each of the relevant terms of the sum, while the
magnitudes of the successive terms are close to each other. One can say that
there is strong destructive interference of ``quantum states" of the system. In
this situation, virtually \textit{any} approximation made in calculating the
individual terms of the sum may alter the precise balance between the terms and
cause large errors in the region of $n \lesssim \bar{N}$. The same problem
appears in our stationary WKB theory, see Section III, which makes the
time-dependent WKB approximation advantageous in this case.

\begin{figure}[h]
\includegraphics[width=8cm,clip=]{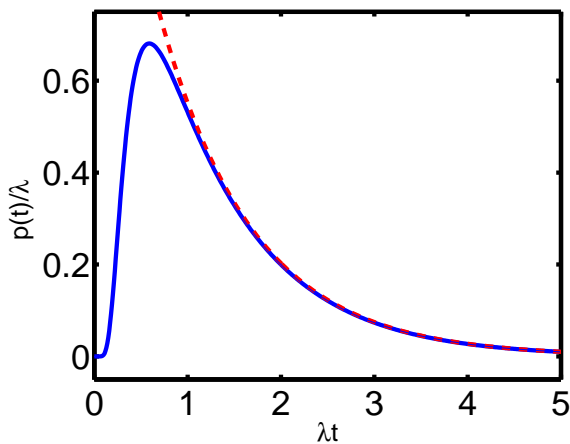}
\caption{(Color online) The lifetime probability distribution $p(t)$
normalized to $\lambda$, as described by Eq. (\ref{ltdistapprox})
(the blue solid line), and its long time asymptote $p(t)/\lambda =
(3/2)\, e^{-\lambda t}$ (the red dashed line).} \label{unidist}
\end{figure}

Now we proceed to calculating the probability distribution of lifetimes of the
particles, and the average extinction time. The quantity $P_0(t)$ is the
probability of extinction at time $t$. Therefore, the lifetime probability
distribution is $p(t)=dP_0(t)/dt$. On the other hand, $P_0(t)=G(x=0,t)$.
Therefore, using Eq. (\ref{gensoleven}), we obtain the exact result
\begin{equation}
\label{ltdist} p(t)= \lambda \sum_{k=1}^{k_0}k(1-2k)\,A_{k}q_{2k}(0)e^{-k(2k -
1)\lambda t}\,.
\end{equation}
Both $p(t)$ and all its derivatives with respect to $t$ vanish  at $t=0$, so
$p(t)$ is exponentially small at $\lambda t \ll 1$. When $k_0\gg 1$ and $\lambda
t\gtrsim 1/\sqrt{k_0}$, Eq. (\ref{ltdist}) can be approximated as
\begin{equation}
\label{ltdistapprox} p(t)\simeq \lambda \sum_{k=1}^{\infty}k(2k -
1)\,\left[{\cal P}_{2k-2}(0)-{\cal P}_{2k}(0)\right] e^{-k (2k-1) \lambda t}\,,
\end{equation}
which is independent of $k_0$. This universal distribution is shown in Fig.
\ref{unidist}. The long-time tail  of the distribution, $\lambda t\gtrsim 1$, is
described by the first term of the sum in Eq. (\ref{ltdistapprox}), which yields
$p(t) \simeq (3 \lambda/2)\, e^{-\lambda t}$.

The average extinction time is
\begin{eqnarray}
\nonumber
\bar{\tau} &=& \int_0^{\infty}t p(t)\, dt = \int_0^{\infty}\left[1-P_0(t) \right]\,dt =\\
&=&\int_0^{\infty}\left[1-G(x=0,t) \right] \,dt\,,
\end{eqnarray}
which yields
\begin{equation}\label{exitime}
\bar{\tau} = \frac{1}{\lambda} \,\sum_{k=1}^{k_0}\frac{A_{k}q_{2k}(0)}{k(2k -
1)}\,.
\end{equation}
When $k_0\gg 1$, we obtain a $k_0$-independent asymptotics
\begin{equation}\label{exitimeasymp}
\bar{\tau} \simeq \frac{1}{\lambda} \,\sum_{k=1}^{\infty}\frac{{\cal
P}_{2k}(0)-{\cal P}_{2k-2}(0)}{k(2k - 1)} = \frac{1.38629 \dots}{\lambda}\,,
\end{equation}
which also follows from Eq. (\ref{ltdistapprox}). Note that the first term in
the series of Eq. (\ref{exitimeasymp}) already gives a fair accuracy:
$\bar{\tau}_1 =1.5/\lambda$.


\begin{thebibliography}{99}
\bibitem{second} The second boundary condition $G(-1,t)=C$ is
specific to the problem in question and results from an additional conservation
law in the binary annihilation reaction. Indeed, in view of Eq.
(\ref{generatingfun}), $G(x=-1,t)=\sum_{m=0}^{\infty} P_{2m}(t) -
\sum_{m=0}^{\infty} P_{2m+1}(t)=C$, while Eq. (\ref{universalBC}) yields
$G(x=1,t)=\sum_{m=0}^{\infty} P_{2m}(t) + \sum_{m=0}^{\infty} P_{2m+1}(t)=1$.
Therefore, the sum of the probabilities of the even states and the sum of the
probabilities of the odd states are conserved separately.
\end{thebibliography}
\end{document}